\documentclass[twocolumn,prb]{revtex4}
\usepackage{graphicx}
\usepackage{dcolumn}
\usepackage{bm}

\begin{document}

\preprint{}
\title{Theory of charge transport in diffusive normal metal
/ conventional superconductor point contacts }
\author{Y. Tanaka$^1$, A.A. Golubov $^2$, and S. Kashiwaya$^3$ }
\affiliation{$^1$Department of Applied Physics, Nagoya University,
Nagoya, 464-8603, Japan \\
$^2$ Faculty of Science and Technology, University of Twente, 7500 AE,
Enschede, The Netherlands \\
$^3$National Institute of Advanced Industrial Science and Technology,
Tsukuba, 305-8568, Japan }
\date{\today}

\begin{abstract}
Tunneling conductance in diffusive normal metal / insulator / s-wave
superconductor (DN/I/S) junctions is calculated for various situations by
changing the magnitudes of the resistance and Thouless energy in DN and the
transparency of the insulating barrier. The generalized boundary condition
introduced by Nazarov [Superlattices and Microstructures \textbf{25}, 1221
(1999)] is applied, where the ballistic theory by Blonder, Tinkham and
Klapwijk (BTK) and the diffusive theory by Volkov, Zaitsev and Klapwijk
based on the boundary condition of Kupriyanov and Lukichev (KL) are
naturally reproduced. It is shown that the proximity effect can enhance
(reduce) the tunneling conductance for junctions with a low (high)
transparency. A wide variety of dependencies of tunneling conductance on
voltage bias is demonstrated including a $U$-shaped gap like structure, a
zero bias conductance peak (ZBCP) and a zero bias conductance dip (ZBCD).
The temperature dependence of tunneling conductance is also calculated and
the conditions for the reentrance effect are studied.
\end{abstract}

\maketitle



%

%




\section{Introduction}


The electron coherence in mesoscopic superconducting systems is one of the
important topics of solid state physics. The low energy transport in these
systems is essentially influenced by the Andreev reflection \cite{Andreev},
a unique process specific for normal metal/superconductor interfaces. The
phase coherence between incoming electrons and Andreev reflected holes
persists in the diffusive normal metal at a mesoscopic length scale and
results in strong interference effects on the probability of Andreev
reflection \cite{Hekking}. These effects become prominent at sufficiently
low temperatures where the thermal broadening is negligible. One of the
remarkable experimental manifestations is the zero bias conductance peak
(ZBCP) \cite%
{Giazotto,Klapwijk,Kastalsky,Nguyen,Wees,Nitta,Bakker,Xiong,Magnee,Kutch,Poirier}%
. A calculation of tunneling conductance in a normal metal (N) /
superconductor (S) junction is an interesting theoretical problem since
quantum interference effects due to Andreev reflection are expected.

For a clean NS contact in the presence of the interface potential barrier
the conductance was calculated by Blonder, Tinkham and Klapwijk \cite{BTK}
(BTK) in terms of the corresponding transmission coefficients on the basis
of the solution of the Bogoljubov - de Gennes equations.
From the general set of boundary conditions connecting the quasiclassical
Green's functions on both sides of the interface for arbitrary transmission
probabilities, Zaitsev \cite{Zaitsev} derived the expression for the
conductance similar to that by BTK. The BTK method \cite{BTK} is confined to
ballistic systems. The generalization of this method to systems with
impurities has been performed by several authors (see the review \cite%
{Beenakker1}). In a number of papers the transmission coefficients were
directly calculated by numerical methods \cite{Lambert,Takane}. However, it
is difficult to apply such methods to most of relevant experimental
situations. Another approach, the so-called random matrix theory, was
employed by Beenakker $et$ $al.$, where the total transmission coefficients
are expressed in terms of those through the normal part of the system and
the normal/superconductor interface separately \cite{Beenakker1,Beenakker2}.
Within this theory, the ZBCP observed in experiments is understood as a
resonance phenomenon related to reflectionless tunneling \cite{reflec}. The
scattering matrix approach was later generalized to finite voltage and
temperature.\cite{Lesovik}

On the other hand, a quasiclassical Green's function calculation based on
nonequilibrium superconductivity theories \cite{Larkin} is much more
powerful and convenient for the actual calculations \cite{Volkov}. In this
approach, the impurity scattering is included in the self-consistent Born
approximation and the weak localization effects are neglected. In the theory
of tunneling conductance developed by Volkov, Zaitsev and Klapwijk (VZK) by
solving the Usadel equations \cite{Usadel}, the origin of the ZBCP observed
in several experiments was clarified to be due to the enhancement of the
pair amplitude in the diffusive normal metal by the proximity effect \cite%
{Volkov}. VZK applied the Kupriyanov and Lukichev (KL) boundary condition
for the Keldysh-Nambu Green's function \cite{KL}. The KL boundary condition
is valid for the atomically sharp interface barrier dividing two diffusive
metals. As shown by Lambert \textit{et al}. \cite{Lambert1}, this condition
is exact in two limits of either high or low barrier transparency, with
small corrections in the intermediate transparency range. By applying the
VZK theory, several authors studied the charge transport in various
junctions \cite%
{Nazarov1,Yip,Stoof,Reentrance,Golubov,Takayanagi,Bezuglyi,Seviour} by
solving the Usadel equations. (see review by Belzig \textit{et al}. \cite%
{Belzig}).

%
%
The generalization of the KL boundary conditions for an arbitrary connector
between diffusive metals was provided by Nazarov 
within the so-called "generalized circuit theory" \cite{Nazarov2}. In this
theory, the mesoscopic system is presented as a network of nodes and
connectors. A connector is characterized by a set of transmission
coefficients and can present anything from a ballistic point contact to a
tunnel junction. A conservation law of matrix current holds in each node.
The method to derive the relation between matrix current and Green's
functions puts the results of Ref.\cite{Zaitsev} to the framework of
Landauer-B\"{u}ttiker scattering formalism.
The boundary condition for Keldysh-Nambu Green's function was derived in
\cite{Nazarov2} for an arbitrary connector including various situations from
ballistic point contact to diffusive contact. Actually, this boundary
condition is very general since the BTK theory is reproduced in the
ballistic limit while in the diffusive limit with a low transmissivity of
the interface, the KL boundary condition is reproduced.

Although a number of papers were published on charge transport in mesoscopic
NS junctions, as far as we know, almost all of them are either based on the
KL boundary conditions or on the BTK model. However in the actual junctions,
transparency of the junction is not necessarily small and impurity
scattering in the DN is important. Therefore, an interesting and important
theoretical problem is the calculation of the tunneling conductance in
normal metal / conventional superconductor junctions using the boundary
condition from Ref.\cite{Nazarov2} since both the ballistic (BTK theory) and
diffusive (VZK theory) cases can be covered simultaneously. In the present
paper, we study the tunneling conductance in diffusive normal metal /
insulator / conventional superconductor (DN/I/S) junctions for various
parameters such as the height of the insulating barrier at the interface,
resistance $R_{d}$ in DN and the Thouless energy $E_{Th}$ in DN. We
concentrate on the normalized tunneling conductance of the junctions $\sigma
_{T}(eV)$ as a function of the bias voltage $V$. The conductance $\sigma
_{T}(eV)$ is given by $\sigma _{T}(eV)=\sigma _{S}(eV)/\sigma _{N}(eV)$
where $\sigma _{S(N)}(eV)$ is the tunneling conductance in the
superconducting (normal ) state at a bias voltage $V$.

In the present paper the following points are clarified:

\noindent 1. When the transparency of the junction is sufficiently low, $%
\sigma _{T}(eV)$ for $\mid eV\mid <\Delta _{0}$ is enhanced with the
increase of $R_{d}$ due to the enhancement of the proximity effect. The ZBCP
becomes prominent for $E_{Th}<<\Delta _{0}$ and $R_{d}/R_{b}<1$. In such a
case, with a further increase of $R_{d}/R_{b}$ the ZBCP changes into a zero
bias conductance dip (ZBCD). In the low transparent limit, the line shapes
of $\sigma _{T}(eV)$ are qualitatively the same as those obtained by VZK
theory \cite{Volkov,Yip}.

\noindent 2. When the transparency of the junction is almost unity, $\sigma
_{T}(eV)$ always exhibits a ZBCD except for the special case of $R_{d}=0$,
i.e. the BTK limit.

\noindent 3. The measure of the proximity effect, $\theta $, is mainly
determined by $R_{d}/R_{b}$ and $E_{Th}$, where $R_{b}$ is the resistance
from the insulating barrier. The proximity effect enhances (reduces) the
magnitude of $\sigma _{T}(eV)$ for junctions with low (high) transparency.

\noindent 4. Even for junctions between conventional $s$-wave
superconductors, we can expect a wide variety of line shapes of the
tunneling conductance, a ZBCP, ZBCD, $U$-shaped structure, and a rounded
bottom structure.

\noindent 5. We have clarified\ the parameter space where a ZBCP should be
expected. Typically, small Thouless energy $E_{Th}$ is required for a ZBCP.
If the magnitude of $E_{Th}$ is increasing up to $\Delta _{0}$, a ZBCP is
only expected for junctions with low transmissivity, $R_{d}/R_{b}<<1$.

The organization of this paper is as follows. In section 2, we will provide
the detailed derivation of the expression for the normalized tunneling
conductance. In section 3, the results of calculations of $\sigma _{T}(eV)$
and $\theta $ are presented for various types of junctions. In section 4,
the summary of the obtained results is given.

\section{Formulation}

In this section we introduce the model and the formalism. We
consider a junction consisting of normal and superconducting
reservoirs connected by a quasi-one-dimensional diffusive
conductor (DN) with a length $L$ much larger than the mean free
path. The interface between the DN\ conductor and the S electrode
has a resistance $R_{d}$ while the DN/N interface has zero
resistance. The positions of the DN/N interface of the DN/S
interface are denoted as $x=0$ and $x=L$, respectively. According
to the circuit theory, the interface between DN and S is
subdivided into two isotropization zones in DN and S, two
ballistic zones and a scattering zone. The sizes of the ballistic
and scattering zones in the current flow direction are much
shorter than the coherence length.
Although the generalized boundary condition of
Ref.~\cite{Nazarov2} is valid for arbitrary interfaces, here
scattering zone is modelled as an infinitely narrow insulating
barrier described by the delta function $U(x)=H\delta(x-L)$.
The resulting  transparency of the junctions $T_{m}$ is given by
$T_{m}=4\cos ^{2}\phi /(4\cos ^{2}\phi +Z^{2})$, where
$Z=2H/(\hbar v_{F})$ is a dimensionless constant, $\phi $ is the
injection angle measured from the interface normal to the junction
and $v_{F}$ is Fermi velocity. Variation of the barrier shape will
not change our results in the considered case of isotropic
superconductivity in the S electrode.

We apply the quasiclassical Keldysh formalism in the following
calculation of the tunneling conductance. The 4 $\times $ 4
Green's functions in DN and S are denoted by $\check{G}_{1}(x)$
and $\check{G}_{2}(x)$ which are expressed in matrix form as

\begin{eqnarray}
\check{G}_{1}(x) &=&\left(
\begin{array}{cc}
\hat{R}_{1}(x) & \hat{K}_{1}(x) \\
0 & \hat{A}_{1}(x)%
\end{array}%
\right) , \\
\check{G}_{2}(x) &=&\left(
\begin{array}{cc}
\hat{R}_{2}(x) & \hat{K}_{2}(x) \\
0 & \hat{A}_{2}(x)%
\end{array}%
\right) ,
\end{eqnarray}%
where the Keldysh component $\hat{K}_{1,2}(x)$ is given by $\hat{K}%
_{1(2)}(x)=\hat{R}_{1(2)}(x)\hat{f}_{1(2)}(x)-\hat{f}_{1(2)}(x)\hat{A}%
_{1(2)}(x)$ with retarded component $\hat{R}_{1,2}(x)$, advanced component $%
\hat{A}_{1,2}(x)$ using distribution function $\hat{f}_{1(2)}(x)$. In the
above, $\hat{R}_{2}(x)$ is expressed by
\[
\hat{R}_{2}(x)=(g\hat{\tau}_{3}+f\hat{\tau}_{2})
\]%
with $g=\epsilon /\sqrt{\epsilon ^{2}-\Delta _{0}^{2}}$ and $f=\Delta _{0}/%
\sqrt{\Delta _{0}^{2}-\epsilon ^{2}}$, where $\epsilon $ denotes the
quasiparticle energy measured from the Fermi energy, $\hat{A}_{2}(x)=-\hat{R}%
_{2}^{\ast }(x)$ and $\hat{f}_{2}(x)=\mathrm{{tanh}[\epsilon /(2k_{B}T)]}$
in thermal equilibrium with temperature $T$.
We put the electrical potential zero in the S-electrode. In this case the
spatial dependence of $\check{G}_{1}(x)$ in DN is determined by the static
Usadel equation \cite{Usadel},

\begin{equation}
D \frac{\partial }{\partial x} [\check{G}_{1} (x) \frac{\partial \check{G}%
_{1}(x) }{\partial x} ] + i [\check{H},\check{G}_{1}(x)] =0,
\end{equation}
with the diffusion constant $D$ in DN, where $\check{H}$ is given by
\[
\check{H}= \left(
\begin{array}{cc}
\hat{H}_{0} & 0 \\
0 & \hat{H}_{0}%
\end{array}
\right),
\]
with $\hat{H}_{0}=\epsilon \tau_{3}. $

The boundary condition for $\check{ G}_{1}(x)$ at the DN/S interface is
given by Nazarov's generalized boundary condition,
\begin{equation}
\frac{L}{R_{d}} (\check{ G}_{1} \frac{ \partial \check{G}_{1} }{\partial x}%
)_{\mid x=L_{-}} =R_{b}^{-1}<B>,  \label{Nazarov}
\end{equation}

\[
B= \frac{ 2T_{m}[\check{G}_{1}(L_{-}),\check{G}_{2}(L_{+})] } { 4 + T_{m}([%
\check{G}_{1}(L_{-}),\check{G}_{2}(L_{+})]_{+} - 2) }.
\]
The average over the various angles of injected particles at the interface
is defined as
\[
<B(\phi)> = \int_{-\pi/2}^{\pi/2} d\phi \cos\phi B(\phi)
/\int_{-\pi/2}^{\pi/2} d\phi T(\phi)\cos\phi
\]
with $B(\phi)=B$ and $T(\phi)=T_{m}$. The resistance of the interface $R_{b}$
is given by
\[
R_{b}=R_{0} \frac{2} {\int_{-\pi/2}^{\pi/2} d\phi T(\phi)\cos\phi}.
\]
Here $R_{0}$ is Sharvin resistance, which in three-dimensional case is given
by $R_{0}^{-1}=e^{2}k_{F}^2S_c/(4\pi^{2}\hbar )$, where $k_{F}$ is the Fermi
wave-vector and $S_c$ is the constriction area. Note that the area $S_c$ is
in general not equal to the crossection area $S_d$ of the normal conductor,
therefore $S_c/S_d$ is independent parameter of our theory.

For $T_{m} \rightarrow 0$ in Eq.~(4), the quantity $B$ can be expressed as
\[
B= \frac{T_{m}}{2}[\check{G}_{1},\check{G}_{2}]
\]
and we can reproduce the KL boundary condition. On the other hand, at $x=0$ $%
\check{G}_{1}(0)$ coincides with that in the normal state.

The electric current is expressed using $\check{G}_{1}(x)$ as
\begin{equation}
I_{el}=\frac{-L}{4eR_{d}}\int_{0}^{\infty }d\epsilon \mathrm{Tr}[\tau _{3}(%
\check{G}_{1}(x)\frac{\partial \check{G}_{1}(x)}{\partial x})^{K}],
\end{equation}%
where $(\check{G_{1}}(x)\frac{\partial \check{G_{1}}(x)}{\partial x})^{K}$
denotes the Keldysh component of $(\check{G_{1}}(x)\frac{\partial \check{%
G_{1}}(x)}{\partial x})$. %
In the actual calculation it is convenient to use the standard $\theta$%
-parameterization when function $\hat{R}_{1}(x)$ is expressed as
\begin{equation}
\hat{R}_{1}(x)=\hat{\tau}_{3}\cos \theta (x)+\hat{\tau}_{2}\sin \theta (x).
\end{equation}
The parameter $\theta (x)$ is a measure of the proximity effect in DN.

Functions $\hat{A}_{1}(x)$ and $\hat{K}_{1}(x)$ are expressed as $\hat{ A}
_{1}(x)=-\hat{R}_{1}^{\ast }(x)$ and $\hat{K}_{1}(x)=\hat{R}_{1}(x)\hat{f }
_{1}(x)-\hat{f}_{1}(x)\hat{A}_{1}(x)$ with the distribution function $\hat{ f%
} _{1}(x)$ which is given by $\hat{f}_{1}(x)=f_{l}(x)+\hat{\tau}_{3}f_{t}(x)
$. In the above, $f_{t}(x)$ is the relevant distribution function which
determines the conductance of the junction we are now concentrating on. From
the retarded or advanced component of the Usadel equation, the spatial
dependence of $\theta (x)$ is determined by the following equation
\begin{equation}
D\frac{\partial ^{2}}{\partial x^{2}}\theta (x)+2i\epsilon \sin [\theta
(x)]=0,  \label{Usa1}
\end{equation}%
%
%
while for the Keldysh component we obtain
\begin{equation}
D\frac{\partial }{\partial x}[\frac{\partial f_{t}(x)}{\partial x} \mathrm{%
cosh^{2}}\theta _{imag}(x)]=0.  \label{Usa2}
\end{equation}%
%
%
At $x=0$, since DN is attached to the normal electrode, $\theta (0)$=0 and $%
f_{t}(0)=f_{t0}$ is satisfied with
\[
f_{t0}=\frac{1}{2}\{\tanh [(\epsilon +eV)/(2k_{B}T)]-\tanh [(\epsilon
-eV)/(2k_{B}T)]\}.
\]%
%
%
Next we focus on the boundary condition at the DN/S interface. Taking the
retarded part of Eq.~(\ref{Nazarov}), we obtain
\begin{equation}
\frac{L}{R_{d}}\frac{\partial \theta (x)}{\partial x}\mid _{x=L_{-}}=\frac{%
<F>}{R_{b}},  \label{b1}
\end{equation}

\[
F = \frac{2 (f \cos\theta_{L} - g \sin\theta_{L})T_{m} } { (2-T_{m}) +
T_{m}[g \cos \theta_{L} +f \sin \theta_{L} ] },
\]
with $\theta_{L}=\theta(L_{-})$.

On the other hand, from the Keldysh part of Eq.~(\ref{Nazarov}), we obtain
\begin{equation}
\frac{L}{R_{d}} (\frac{\partial f_{t}}{ \partial x}) \mathrm{{\cosh^{2}}}
\theta_{imag}(x) \mid_{x=L_{-}} =-\frac{<I_{b0}> f_{t}(L_{-})}{R_{b}},
\label{b2}
\end{equation}
with
\[
I_{b0} = \frac{T_{m}^{2} \Lambda_{1} + 2T_{m}(2-T_{m}) \Lambda_{2}} {2 \mid
(2-T_{m}) + T_{m}[g \cos\theta_{L} + f \sin\theta_{L} ] \mid^{2} },
\]
\[
\Lambda_{1}=(1+\mid \cos\theta_{L} \mid^{2} + \mid \sin\theta_{L} \mid^{2})
(\mid g \mid^{2} + \mid f \mid^{2} +1)
\]
\begin{equation}
+ 4\mathrm{Imag}[fg^{*}] \mathrm{Imag}[\cos \theta_{L} \sin\theta_{L}^{*} ],
\end{equation}

\begin{equation}
\Lambda_{2} =\mathrm{{Real} \{ g(\cos \theta_{L} + \cos \theta_{L}^{*}) +
f(\sin \theta_{L} + \sin \theta_{L}^{*}) \}},
\end{equation}
where $\theta_{imag}(x)$ denotes the imaginary part of
$\theta(x)$.
For $T_{m}<<1$, $I_{b0}$ is reduced to
\begin{equation}
I_{b0}=\frac{\mathrm{{\ Real}[g(\cos \theta _{L}+\cos \theta _{L}^{\ast
})+f(\sin \theta _{L}+\sin \theta _{L}^{\ast })]}}{2}T_{m},
\end{equation}%
which is the expression used in the VZK theory.
After a simple manipulation, we can obtain $f_{t}(L_{-})$
\[
\displaystyle f_{t}(L_{-})=\frac{R_{b}f_{t0}}{R_{b}+\frac{R_{d}<I_{b0}>}{L}%
\int_{0}^{L}\frac{dx}{\cosh ^{2}\theta _{imag}(x)}}
\]%
Since the electric current $I_{el}$ can be expressed via $\theta _{L}$ in
the following form%
\[
I_{el}=-\frac{L}{eR_{d}}\int_{0}^{\infty }(\frac{\partial f_{t}}{\partial x}%
)\mid _{x=L_{-}}\cosh ^{2}[\mathrm{Imag}(\theta _{L})]d\epsilon ,
\]%
%
%
we obtain the following final result for the current

\begin{equation}
I_{el}=\frac{1}{e}\int_{0}^{\infty }d\epsilon \frac{f_{t0}}{\frac{R_{b}}{%
<I_{b0}>}+\frac{R_{d}}{L}\int_{0}^{L}\frac{dx}{\cosh ^{2}\theta _{imag}(x)}}.
\end{equation}%
Then the total resistance $R$ at zero temperature is given by
\begin{equation}
R=\frac{R_{b}}{<I_{b0}>}+\frac{R_{d}}{L}\int_{0}^{L}\frac{dx}{\cosh
^{2}\theta _{imag}(x)}
\end{equation}
and the tunneling conductance in the superconducting state $\sigma _{S}(eV)$
is given by $\sigma _{S}(eV)=1/R$.

It should be mentioned that for $R_{d}=0$, $\theta_{L}$ becomes zero due to
the absence of the proximity effect. Then $I_{b0}$ is given as follows
\[
I_{b0}=\frac{(1+\mid g\mid ^{2}+\mid f\mid ^{2}) T_{m}^{2}+2T_{m}(2-T_{m})%
\mathrm{Real}(g)}{\mid (2-T_{m})+T_{m}g \mid^{2}}
\]%
\begin{equation}
=\frac{T_{m}[1+\mid \Gamma \mid ^{2}+(T_{m}-1)\mid \Gamma \mid ^{4}]}{\mid
1-(1-T_{m})\Gamma ^{2}\mid ^{2}}
\end{equation}%
with $\Gamma =\frac{\sqrt{\epsilon -\sqrt{\epsilon ^{2}-\Delta _{0}^{2}}}}{%
\sqrt{\epsilon +\sqrt{\epsilon ^{2}-\Delta _{0}^{2}}}}$ and the resulting $%
\sigma _{S}$ is given by
\[
\sigma_{S}(eV)=\frac{1}{R_{0}} \int^{\pi/2}_{-\pi/2} \frac{I_{b0}}{2} \cos
\phi d\phi,
\]
and reproduces that by BTK theory.

It should be remarked that in the present circuit theory, $R_{d}/R_{b}$ can
be varied independently of $T_{m}$, $i.e.$ independently
of $Z$,
since one
can change the magnitude of the constriction area $S_c$ independently. In
other words, $R_{d}/R_{b}$ is no more proportional to $T_{av}(L/l)$, where $%
T_{av}$ is the averaged transmissivity of the barrier and $l$ is the mean
free path in the diffusive region, respectively. Based on this fact, we can
choose $R_{d}/R_{b}$ and $Z$ as independent parameters.

%
In the following section, we will discuss the normalized tunneling
conductance $\sigma _{T}(eV)=\sigma _{S}(eV)/\sigma _{N}(eV)$ where $\sigma
_{N}(eV)$ is the tunneling conductance in the normal state given by $\sigma
_{N}(eV)=\sigma _{N}=1/(R_{d}+R_{b})$, respectively.


\section{Results}

\subsection{Tunneling conductance vs voltage: zero-bias anomalies}

In this section, we focus on the line shape of the tunneling conductance.
Let us first choose the relatively strong barrier $Z=10$ (Figs. 1 and 2) for
various $R_{d}/R_{b}$. For $E_{Th}=\Delta _{0}$, the magnitude of $\sigma
_{T}(eV)$ for $\mid eV\mid <\Delta _{0}$ increases with the increase of $%
R_{d}/R_{b}$. First, the line shape of the tunneling conductance remains to
be $U$ shaped and only the height of the bottom value is enhanced (curve $b$
). Then, with a further increase of $R_{d}/R_{b}$, a rounded bottom
structure (curve $c$ and $d$) appears and finally it changes into a nearly
flat line shape (curve $e$). 
\begin{figure}[bh]
\begin{center}
\scalebox{0.4}{
\includegraphics[width=10.0cm,clip]{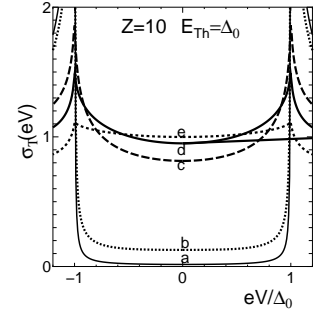}}
\end{center}
\caption{ Normalized tunneling conductance for Z=10 and $E_{Th}/\Delta
_{0}=1 $. a: $R_{d}/R_{b}=0$, b: $R_{d}/R_{b}=0.1$, c: $R_{d}/R_{b}=1$, d: $%
R_{d}/R_{b}=2$ and e: $R_{d}/R_{b}=10$. }
\end{figure}
For $E_{Th}=0.01\Delta _{0}$ (Fig. 2), the magnitude of $\sigma _{T}(eV)$
has a ZBCP once the magnitude of $R_{d}/R_{b}$ deviates slightly from 0. The
order of magnitude of the ZBCP width is given by $E_{Th}$. When the
magnitude of $R_{d}/R_{b}$ exceeds unity, the ZBCP splits into two (curve $d$%
) and finally $\sigma _{T}(eV)$ acquires a zero bias conductance dip (ZBCD)
(curve $e$). 
In the limit of an extremely strong barrier with $Z\rightarrow \infty $, the
results of the VZK theory are reproduced.

\begin{figure}[bh]
\begin{center}
\scalebox{0.4}{
\includegraphics[width=10.0cm,clip]{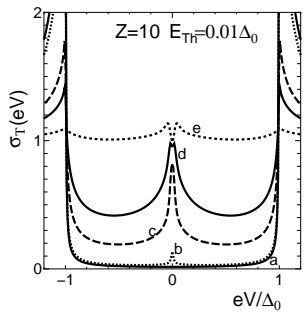}}
\end{center}
\caption{ Normalized tunneling conductance for Z=10 and $E_{Th}/\Delta
_{0}=0.01$. a: $R_{d}/R_{b}=0$, b: $R_{d}/R_{b}=0.1$, c: $R_{d}/R_{b}=1$, d:
$R_{d}/R_{b}=2$ e: $R_{d}/R_{b}=10$. }
\label{fig:01B}
\end{figure}
%
On the other hand, in the fully transparent case with $Z=0$ the line shape
of the tunneling conductance becomes quite different. For $E_{Th}=\Delta
_{0} $ (Fig. 3), the magnitude of $\sigma _{T}(eV)$ decreases with the
increase of the magnitude of $R_{d}/R_{b}$. The bottom parts of all curves
are rounded and $\sigma _{T}(eV)$ always exceeds unity. On the other hand,
for $E_{Th}=0.01\Delta _{0}$, $\sigma _{T}(eV)$ has a ZBCD even for a small
magnitude of $R_{d}/R_{b}$ except for the special case of $R_{d}/R_{b}=0$
where the BTK theory is valid. This feature is quite different from that
shown in Fig. 2. %
\begin{figure}[tbh]
\begin{center}
\scalebox{0.4}{
\includegraphics[width=10.0cm,clip]{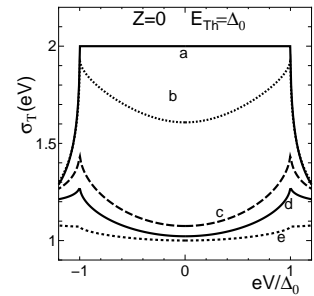}}
\end{center}
\caption{ Normalized tunneling conductance for $Z=0$ and $E_{Th}/\Delta
_{0}=1$. a: $R_{d}/R_{b}=0$, b: $R_{d}/R_{b}=0.1$, c: $R_{d}/R_{b}=1$ d: $%
R_{d}/R_{b}=2$ and e: $R_{d}/R_{b}=10$. }
\label{fig:04A}
\end{figure}
%
%
\begin{figure}[tbh]
\begin{center}
\scalebox{0.4}{
\includegraphics[width=10.0cm,clip]{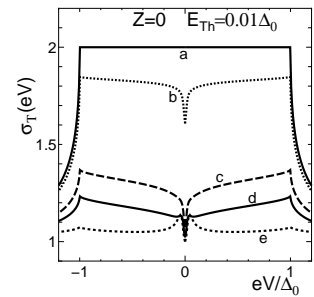}}
\end{center}
\caption{ Normalized tunneling conductance for $Z=0$ and $E_{Th}/\Delta
_{0}=0.01$. a: $R_{d}/R_{b}=0$, b: $R_{d}/R_{b}=0.1$, c: $R_{d}/R_{b}=1$ d: $%
R_{d}/R_{b}=2$ and e: $R_{d}/R_{b}=10$. }
\label{fig:04B}
\end{figure}
In the case of an intermediate barrier strength, $Z=1$, the shape of $\sigma
_{T}(eV)$ becomes rather complex. For $E_{Th}=\Delta _{0}$, $\sigma _{T}(eV)$
has a shallow gap structure similar to the case of the BTK theory (curve $a$
). With the increase of $R_{d}/R_{b}$, the coherent peak structure at $%
eV=\pm \Delta _{0}$ is smeared out and the voltage dependence becomes very
weak as shown by curve $e$. For $E_{Th}=0.01\Delta _{0}$, the ZBCP appears
for a small magnitude of $R_{d}/R_{b}$ (see curve $b$). With increasing
magnitude of $R_{d}/R_{b}$, the ZBCP changes into a ZBCD (curves $c$ $d$ $e$%
). As compared to the $Z=10$ case (see Fig. 2), the ZBCP is much more easier
to change into a ZBCD by increasing the magnitude of $R_{d}/R_{b}$.
\begin{figure}[tbh]
\begin{center}
\scalebox{0.4}{
\includegraphics[width=10.0cm,clip]{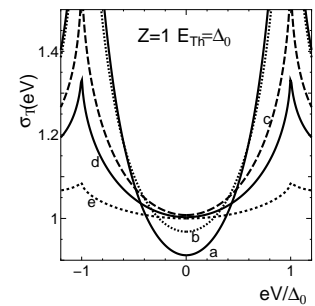}}
\end{center}
\caption{ Normalized tunneling conductance for $Z=1$ and $E_{Th}/\Delta
_{0}=1$. a: $R_{d}/R_{b}=0$, b: $R_{d}/R_{b}=0.1$, c: $R_{d}/R_{b}=1$ d: $%
R_{d}/R_{b}=2$ and e: $R_{d}/R_{b}=10$. }
\label{fig:04A}
\end{figure}
%
\begin{figure}[tbh]
\begin{center}
\scalebox{0.4}{
\includegraphics[width=10.0cm,clip]{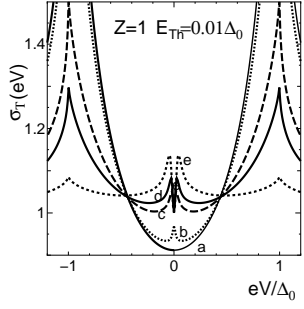}}
\end{center}
\caption{ Normalized tunneling conductance for $Z=1$ and $E_{Th}/\Delta
_{0}=0.01$. a: $R_{d}/R_{b}=0$, b: $R_{d}/R_{b}=0.1$, c: $R_{d}/R_{b}=1$ d: $%
R_{d}/R_{b}=2$ and e: $R_{d}/R_{b}=10$. }
\label{fig:04B}
\end{figure}

It is interesting to study how various parameters influence the proximity
effect. The measure of the proximity effect at the S/N interface $\theta
_{L} $ is plotted for $Z=0$ and $Z=10$ with corresponding parameters in
Figs. 1 to 4. For $R_{d}/R_{b}=0$, $\theta _{L}=0$ is satisfied for any $%
E_{Th}$ and $Z$. Besides this fact, at $\epsilon =0$, $\theta _{L}$ always
becomes a real number. First, we study the case of $E_{Th}/\Delta _{0}=1$
(Fig. 7) where the same values of $R_{d}/R_{b}$ are chosen as in Figs. 1 and
3. The real part of $\theta _{L}$ is enhanced with an increase in $%
R_{d}/R_{b}$ and is almost constant as function of $\epsilon $. At the same
time, the imaginary part of $\theta _{L}$ is an increasing function of $%
\epsilon $ for all cases. There is no clear qualitative difference between
the energy dependencies of $\mathrm{Real}(\theta _{L})$ and $\mathrm{Imag}%
(\theta _{L})$ at $Z=0$ and $Z=10$.
\begin{figure}[tbh]
\begin{center}
\scalebox{0.45}{
\includegraphics[width=12.0cm,clip]{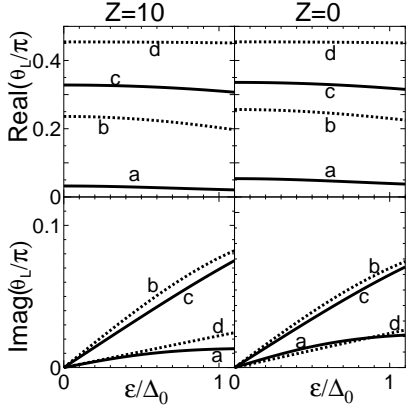}}
\end{center}
\caption{ Real part of $\protect\theta _{L}$ (upper panels) and imaginary
part of it (lower panels) is plotted as a function of $\protect\epsilon $.
Z=10 (left panels) and Z=0 (right panels) with $E_{Th}/\Delta _{0}=1$. a: $%
R_{d}/R_{b}=0.1$ b: $R_{d}/R_{b}=1$, c: $R_{d}/R_{b}=2$ and d: $%
R_{d}/R_{b}=10$. }
\label{fig:06}
\end{figure}
%
%
Next, we discuss the line shapes of $\theta _{L}$ for $E_{Th}/\Delta
_{0}=0.01$. Real($\theta _{L}$) has a peak at zero voltage and decreases
with the increase of $\epsilon $. Imag($\theta _{L}$) increases sharply from
0 and has a peak at about $\epsilon \sim E_{Th}$, except for a sufficiently
large value of $R_{d}$. Also in this case, there is no qualitative
difference between the line shapes of $\mathrm{Real(Imag)}(\theta _{L})$ for
the $Z=0$ case and that for $Z=10$.

%
\begin{figure}[tbh]
\begin{center}
\scalebox{0.45}{
\includegraphics[width=12.0cm,clip]{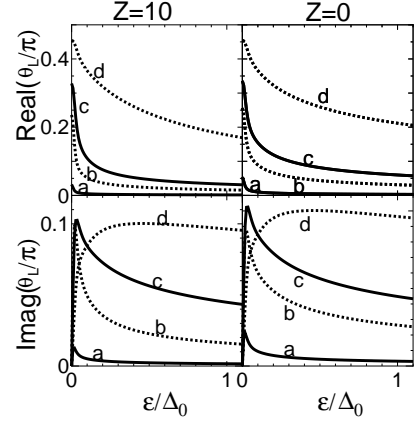}}
\end{center}
\caption{ Real part of $\protect\theta _{L}$ (upper panels) and imaginary
part of it (lower panels) is plotted as a function of $\protect\epsilon $.
Z=10 (left panels) and Z=0 (right panels) with $E_{Th}/\Delta _{0}=0.01$. a:
$R_{d}/R_{b}=0.1$ b: $R_{d}/R_{b}=1$, c: $R_{d}/R_{b}=2$ and d: $%
R_{d}/R_{b}=10$. }
\label{fig:07}
\end{figure}
Although the magnitude of $\theta _{L}$, $i.e.$ the measure of proximity
effect, is enhanced with increasing $R_{d}/R_{b}$, its influence on $\sigma
_{T}(eV)$ is different for low and high transparent junctions. In the low
transparent junctions, the increase in the magnitude of $\theta _{L}$ by $%
R_{d}/R_{b}$ can enhance the conductance $\sigma _{T}(eV)$ for $eV\sim 0$
and produce a ZBCP, whereas in high transparent junctions the enhancement of
$\theta _{L}$ suppresses the magnitude of $\sigma _{T}(eV)$.

Finally, we focus on the condition where the ZBCP appears. We change $Z$ and
$R_{d}/R_{b}$ for fixed $E_{Th}$. An upper critical value of $R_{d}=R_{bu}$
exists where the ZBCP vanishes for $R_{d}>R_{bu}$. For $E_{Th}=0.01\Delta
_{0}$, $R_{bu}$ increases with $Z$ and converges at nearly 1.4. For $%
E_{Th}=0.8\Delta _{0}$, the lower critical value of $R_{d}=R_{bl}$ also
appears where the ZBCP vanishes for $R_{d}<R_{bl}$. The ZBCP is expected for
$R_{bl}<R_{d}<R_{bu}$. The magnitude of $R_{bu}$ is suppressed drastically
as compared to that for $E_{Th}=0.01\Delta _{0}$. For $E_{Th}>\Delta _{0}$,
a ZBCP region vanishes for $0<Z<20$. %
\begin{figure}[tbh]
\begin{center}
\scalebox{0.4}{
\includegraphics[width=13.0cm,clip]{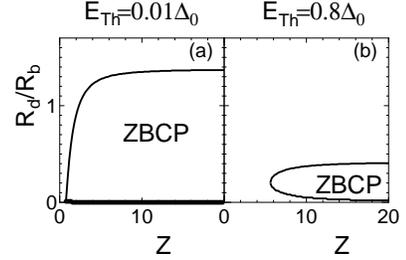}}
\end{center}
\caption{ The parameter space where ZBCP appears. }
\label{fig:09}
\end{figure}
In order to understand the crossover from the ZBCP to the ZBCD in much more
detail, it is interesting to calculate the second derivative of the total
resistance $R$ as a function of $\epsilon =eV$ at $\epsilon =0$. For
simplicity, here we focus on the case of a sufficiently large $Z$. For $%
\epsilon <\Delta _{0}$, for simplicity total resistance is written as
\begin{equation}
R=R_{1}+R_{2},
\end{equation}%
\[
R_{1}=\frac{R_{b}}{<I_{b0}>},\ \ R_{2}=\frac{R_{d}}{L}\int_{0}^{L}\frac{dx}{%
\cosh ^{2}\theta _{i}(x)},
\]%
\[
<I_{b0}>=f\sin \theta _{L,r}\mathrm{cosh}\theta _{L,i},
\]%
with $\theta _{L}=\theta _{L,r}+i\theta _{L,i}$, where $\theta _{L,r}$ and $%
\theta _{L,i}$ are the real and imaginary parts of $\theta _{L}$,
respectively. The second derivative of $I_{b0}$ at $\epsilon =0$ is given by
\[
\frac{\partial ^{2}I_{b0}}{\partial \epsilon ^{2}}=\frac{\sin \theta _{L,r}}{%
\Delta _{0}^{2}}+\cos \theta _{L,r}\frac{\partial ^{2}\theta _{L,r}}{%
\partial \epsilon ^{2}}+\sin \theta _{L,r}(\frac{\partial \theta _{L,i}}{%
\partial \epsilon })^{2},
\]%
since $\frac{\partial \theta _{L,r}}{\partial \epsilon }=0$ and $\theta
_{L,i}=0$ is satisfied at $\epsilon =0$.

The resulting second derivative of the total resistance at zero energy is
given as
\[
\frac{ \partial^{2} R }{\partial \epsilon^{2}} =\frac{ \partial^{2} R_{1} }{
\partial \epsilon^{2}} + \frac{ \partial^{2} R_{2} }{\partial \epsilon^{2}},
\]
\[
\frac{ \partial^{2} R_{1} }{\partial \epsilon^{2}} =
\]
\begin{equation}
-\frac{R_{b} }{\sin^{2} \theta_{L,r} } [\frac{\sin \theta_{L,r}}{%
\Delta_{0}^{2}} + \cos \theta_{L,r} \frac{ \partial^{2} \theta_{L,r} }{%
\partial \epsilon^{2}} + \sin \theta_{L,r} (\frac{\partial \theta_{L,i} }{%
\partial \epsilon})^{2}],  \label{rela}
\end{equation}

\begin{equation}
\frac{\partial ^{2}R_{2}}{\partial \epsilon ^{2}}=-\frac{2R_{d}}{L}%
\int_{0}^{L}[\frac{\partial \theta _{i}(x)}{\partial \epsilon }]^{2}dx,
\end{equation}%
where $\theta _{i}(x)$ is the imaginary part of $\theta (x)$. The sign of $%
\frac{\partial ^{2}\theta _{L,r}}{\partial \epsilon ^{2}}$ becomes negative
and it can induce $\frac{\partial ^{2}R_{1}}{\partial \epsilon ^{2}}>0$ in
some cases. The order of $\frac{\partial ^{2}\theta _{L,r}}{\partial
\epsilon ^{2}}$ and $(\frac{\partial \theta _{L,i}}{\partial \epsilon })^{2}$
is proportional to the inverse of $E_{Th}^{2}$. The sign of $\frac{\partial
^{2}R_{1}}{\partial \epsilon ^{2}}$ is crucially influenced by the relative
magnitude of the second term at the right-hand side of Eq.~(\ref{rela}). On
the other hand, the sign of $\frac{\partial ^{2}R_{2}}{\partial \epsilon ^{2}%
}$ is always negative.

For $E_{Th}<<\Delta _{0}$ and small magnitude of $R_{d}/R_{b}$, the
resulting $\theta _{L,r}$ is sufficiently small and the magnitude of $\frac{%
\partial ^{2}R_{1}}{\partial \epsilon ^{2}}$ becomes positive. When this
positive contribution overcomes the negative contribution from $\frac{%
\partial ^{2}R_{2}}{\partial \epsilon ^{2}}$, we can expect a resistance
minimum at zero energy, $i.e.$, a ZBCP. However, with increasing $R_{d}/R_{b}
$, the magnitude of $\frac{\partial ^{2}R_{1}}{\partial \epsilon ^{2}}$
decreases due to the enhancement of the third term in the right-hand side of
Eq.~(\ref{rela}), while $\frac{\partial ^{2}R_{2}}{\partial \epsilon ^{2}}$
increases. Then, a critical value $R_{d}=R_{bu}$ appears, above which the
ZBCP changes into a ZBCD with an increase of $R_{d}/R_{b}$. This is the
mechanism of the crossover from a ZBCP to a ZBCD.

When the magnitude of $E_{Th}$ is enhanced, the magnitudes of the second and
third terms in Eq.~(\ref{rela}) are reduced and the first term can not be
neglected. The resulting magnitude of $\frac{\partial ^{2}R_{1}}{\partial
\epsilon ^{2}}$ is suppressed and the value of $R_{bu}$ is reduced. This is
the origin of the difference in $R_{bu}$ in Figs. 9(a) and (b).

It is also interesting to present similar plots as Fig. 9 using an averaged
transparency of the junction $T_{av}$
\begin{equation}
T_{av}=\frac{\int_{-\pi /2}^{\pi /2}\cos \phi T(\phi )d\phi }{2}.
\end{equation}%
The results are shown in Fig. 10. For $E_{Th}=0.01\Delta _{0}$, the
magnitude of $R_{bu}$ decreases monotonically with increasing $T_{av}$ and
it vanishes about $T_{av}\sim 0.8$. For $E_{Th}=0.8\Delta _{0}$, the
magnitude of $R_{bl}$ increases for an increasing $T_{av}$ while that for $%
R_{bu}$ decreases. The ZBCP region is restricted to small values of $%
R_{d}/R_{b}$.

%
\begin{figure}[htb]
\begin{center}
\scalebox{0.4}{
\includegraphics[width=13.0cm,clip]{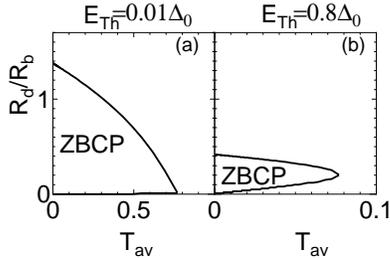}}
\end{center}
\caption{ Similar plots as in Fig. 9 using $T_{av}$. }
\label{fig:10}
\end{figure}

In any way, the preferred condition for the formation of a ZBCP is the
combination of the low transparency of the junction and the smallness of the
$E_{Th}/\Delta _{0}$ ratio. This situation is understood as follows. It is
well known from the BTK theory that the magnitude of the zero bias
conductance is almost proportional to $T_{av}^{2}$ for $R_{d}=0$ for $%
T_{av}<<1$. With the increase in the magnitude of $R_{d}$, the measure of
the proximity effect $\theta _{L}$ is enhanced for $\mid \epsilon \mid
<E_{Th}$ and the zero bias conductance is proportional to $T_{av}$. However,
the magnitude of $\theta _{L}$ at finite energy in the range $E_{Th}<\mid
\epsilon \mid <\Delta _{0}$ is drastically suppressed as shown in Fig. 8,
due to the existence of the proximity induced minigap \cite{minigap} in the
normal diffusive part DN of the order of $E_{Th}$. As a result, in this
regime the conductance channel that provides a contribution proportional to $%
T_{av}$ is not available. Thus only the low voltage conductance is enhanced.
On the other hand, for large $E_{Th}$ with $E_{Th}\sim \Delta _{0}$, the
measure of proximity effect $\theta _{L}$ is insensitive to energy for $\mid
\epsilon \mid <\Delta _{0}$ (see Fig. 7), then the resulting $\sigma
_{T}(eV) $ is always enhanced for $\mid eV\mid <\Delta _{0}$ and the degree
of the prominent enhancement of $\sigma _{T}(0)$ is weakened.

\subsection{Tunneling conductance vs temperature: reentrance effect}

Finally, we look at temperature dependence of conductance for various $Z$
and $R_{d}/R_{b}$ and focus on the relevance to the corresponding results in
VZK theory based on the KL boundary condition. We calculate tunneling
conductance at non zero temperature following
\begin{equation}
\sigma_{S}(eV,T)=dI_{el}/dV.
\end{equation}
Then we define deviation of tunneling conductance from that at zero
temperature given by
\begin{equation}
\delta \sigma_{S} =\sigma_{S}(eV,T)-\sigma_{S}(eV,0).
\end{equation}
In the following, we will plot $\delta \sigma_{S}/\sigma_{N}$ as a function
of temperature $T$.

\begin{figure}[tbh]
\begin{center}
\scalebox{0.4}{
\includegraphics[width=10.0cm,clip]{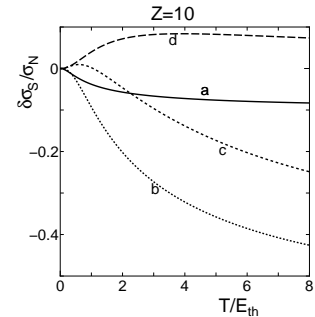}}
\end{center}
\caption{ $\protect\delta \protect\sigma_{S}/\protect\sigma_{N}$ is plotted
as a function of $T$. Z=10 and $E_{Th}=0.01\Delta_{0}$. a: $R_{d}/R_{b}=0.1$
b: $R_{d}/R_{b}=1$, c: $R_{d}/R_{b}=2$ and d: $R_{d}/R_{b}=10$. }
\label{fig:11}
\end{figure}

For $R_{d}/R_{b}=0.1$ (curve $a$) and $R_{d}/R_{b}=1$ (curve $b$), due to
the existence of ZBCP as shown in Fig. 2, $\delta \sigma_{S}/\sigma_{N}$
takes negative value and decreases with $T$. While for $R_{d}/R_{b}=2$, $%
\delta \sigma_{S}/\sigma_{N}$ first increases and decreases again (curve $c$%
). With the further increase of the magnitude of $R_{d}/R_{b}$, $\delta
\sigma_{S}/\sigma_{N}$ becomes positive and it has maximum at a certain
temperature. This effect is known as "reentrance effect". It was predicted
theoretically within the VZK theory in \cite{Stoof,Reentrance,Golubov} and
observed experimentally \cite{Reentrance_exp}. According to the theory \cite%
{Stoof,Golubov,Reentrance}, for $R_{b} \rightarrow 0$, the maximum value of $%
\delta \sigma_{S}/\sigma_{N}$ is about 0.09 at temperature of the
order of Thouless energy. Note that the reentrance of the metallic
conductance occurs as a function of bias voltage as well, but here
we concentrate on the temperature dependence since it was studied
in most detail within the VZK theory.

\begin{figure}[tbh]
\begin{center}
\scalebox{0.4}{
\includegraphics[width=10.0cm,clip]{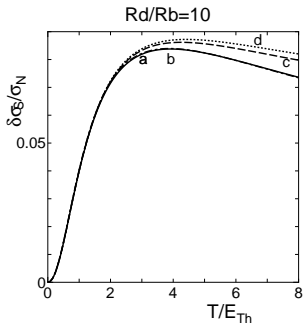}}
\end{center}
\caption{ $\protect\delta \protect\sigma_{S}/\protect\sigma_{N}$ is plotted
as a function of $T$. $R_{d}/R_{b}=10$ and $E_{Th}=0.01\Delta_{0}$. a: VZK
theory, b: $Z=10$, c: $Z=1$ and d: $Z=0$. }
\label{fig:12}
\end{figure}
In order to study the consequences of our theory for the reentrance effect,
we have calculated the temperature dependence of $\delta
\sigma_{S}/\sigma_{N}$ for various $Z$ for fixed $R_{d}/R_{b}$. For the
comparison with the standard VZK theory, we also plot $\delta
\sigma_{S}/\sigma_{N}$ using the KL boundary condition. For $R_{d}/R_{b}=10$%
, we always see the standard reentrant behavior. For $Z=10$, we can not see
clear deviation from the VZK theory (compare curves $a$ and $b$). With the
decrease of the magnitude of $Z$, the magnitude of $\delta
\sigma_{S}/\sigma_{N}$ is enhanced much stronger (curves $c$ and $d$).
Although there are quantitative difference between four curves, the
qualitative line shapes are similar to those predicted by the VZK theory.

\begin{figure}[tbh]
\begin{center}
\scalebox{0.4}{
\includegraphics[width=10.0cm,clip]{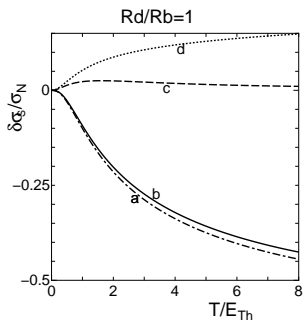}}
\end{center}
\caption{ $\protect\delta \protect\sigma_{S}/\protect\sigma_{N}$ is plotted
as a function of $T$. $R_{d}/R_{b}=1$ and $E_{Th}=0.01\Delta_{0}$. a: VZK
theory, b: $Z=10$, c: $Z=1$ and d: $Z=0$. }
\label{fig:13}
\end{figure}
However, situation is different for decreasing $R_{d}/R_{b}=1$. For $%
R_{d}/R_{b}=1$ the resulting $\delta \sigma_{S}/\sigma_{N}$ takes negative
value and similar feature is also obtained for $Z=10$ (see curves $a$ and $b$%
). At the same time, for small magnitude of $Z$, $\delta
\sigma_{S}/\sigma_{N}$ takes positive value. As seen from these results, the
deviation from VZK becomes significant for small magnitude of $Z$ with $%
R_{d}/R_{b} < 1$.

\section{Conclusions}

In the present paper, a detailed theoretical investigation of the tunneling
conductance of diffusive normal metal / conventional superconductor
junctions is presented. Even for conventional s-wave junctions, the
interplay between diffusive and interface scattering produces a wide variety
of line shapes of the tunneling conductance: ZBCP, ZBCD, U-shaped, and
rounded bottom structures. There are several points which have been
clarified in this paper.

1. When the transparency of a junction is sufficiently low and $E_{Th}$ is
small, the ZBCP appears for $R_{d}<R_{b}$. With an increase in $R_{d}$, the
ZBCP changes into a zero bias conductance dip (ZBCD). For large $E_{Th}$
with the same order of $\Delta _{0}$, the ZBCP is only expected for small
values of $R_{d}$. With increasing $E_{Th}$, the ZBCP vanishes when $%
E_{Th}>\Delta _{0}$ is satisfied. For the low transparency limit, the
results obtained by us are reduced to those by the VZK theory where the KL
boundary condition is used.

2. When the transparency of the junction is almost unity, $\sigma_{T}(eV)$
always have a ZBCD except for the case of vanishing $R_{d}$.

3. The proximity effect can enhance (reduce) the tunneling conductance of
junctions with low (high) transparency.

The above mentioned criteria for the existence of a ZBCP agree with
available experimental data. However, we are not aware of an experimental
observation of ZBCD in highly transparent junctions.

In the present paper, the superconductor is restricted to be a conventional
s-wave superconductor. However, it is well known that a ZBCP also appears in
unconventional superconductor junctions \cite{TK95,Kashi00}, the origin of
which is the formation of midgap Andreev bound states (MABS) \cite{Buch}.
Indeed, a ZBCP has been reported in various superconductors that have an
anisotropic pairing symmetry. \cite{Kashi00,Experiments} It should be
remarked that the line shape of the ZBCP obtained in the present paper is
quite different from that by MABS. In the present case, the height of $%
\sigma _{T}(eV)$ never exceeds unity and its width is determined by $E_{Th}$%
, while in the MABS case, the peak height is proportional to the inverse of
the magnitude of $T_{m}$ and the width is proportional to $T_{m}$.

The proper theory of transport of unconventional junctions in the presence
of MABS has been formulated \cite{TK95,Kashi00} only under the conditions of
ballistic transport. Recently, this theory has been revisited to account for
diffusive transport in the normal metal in Ref.\cite{Nazarov3}, where an
extension of the circuit theory was provided for unconventional
superconductor junctions. A general relation was derived for matrix current,
$B$ in Eq. (4), which is available for unconventional superconductor
junctions with MABS. An elaborated example demonstrated the interplay of
MABS and proximity effect in a $d$-wave junction. It is actually quite
interesting to apply this novel circuit theory to the actual calculation of $%
\sigma _{T}(eV)$ as in the present paper. Such a direction of study is
important in order to analyze recent tunneling experiments where mesoscopic
interference effects were observed in high-T$_{C}$ junction systems \cite%
{Hiromi}. We will present the obtained results in the near future \cite%
{Golubov2}.

In the present study, we have focused on N/S junctions. The extension of
Nazarov's circuit theory to long diffusive S/N/S junctions has been
performed by Bezuglyi \textit{et al} \cite{Bezuglyi}. In S/N/S junctions,
the mechanism of multiple Andreev reflections produces the subharmonic gap
structures on I-V curves \cite{M1,M2,M3,M4,M5,M6,M7,M8} and the situation
becomes much more complex as compared to N/S junctions. Moreover, in S/N/S
junctions with unconventional superconductors, MABS lead to the anomalous
current-phase relation and temperature dependence of the Josephson current
\cite{TKJ}. An interesting problem is an extension of the circuit theory to
S/N/S junctions with unconventional superconductors.

%
The authors appreciate useful and fruitful discussions with Yu. Nazarov, J.
Inoue, and H. Itoh. This work was supported by the Core Research for
Evolutional Science and Technology (CREST) of the Japan Science and
Technology Corporation (JST). The computational aspect of this work has been
performed at the facilities of the Supercomputer Center, Institute for Solid
State Physics, University of Tokyo and the Computer Center.
%


\end{document}